\documentclass[
    ,final            
  ]
  {aipproc}

\layoutstyle{6x9}
\usepackage{amsmath}
\usepackage{graphicx}  

\def\la{\langle}
\def\ra{\rangle}
\def\ph{{ \phi _h}}

\def\cs{\la {\cos\ph} \ra}
\def\cst{\la {\cos2\ph} \ra}
\def\csm{\cos\ph}
\def\cstm{\cos2\ph}

\title{{Measurement of azimuthal asymmetries of the unpolarized 
cross section at HERMES}}
\date{(on behalf of the HERMES Collaboration)}


\author{Francesca Giordano}
{address={INFN \& Universit\`a degli studi di Ferrara, giordano@fe.infn.it}}
\author{Rebecca Lamb} 
{address={University of Illinois, rlamb2@illinois.edu}}

\begin{abstract}
A multi-dimensional ($x, y, z, P_{h\perp}$) 
extraction of $\cos \phi_h$ and $\cos 2\phi_h$ azimuthal asymmetries 
of unpolarized Semi-Inclusive 
Deep Inelastic Scattering at HERMES is discussed. 
The use of data taken with hydrogen and deuterium targets and the separation of positive and 
negative hadrons allow to access flavor-dependent information 
about quark intrinsic transverse momenta and spin-orbit correlations. 
This flavor sensitivity allows for a discrimination between 
theoretical models in the HERMES kinematic regime.
\end{abstract}

\begin{document}
\maketitle

\section{Introduction}\label{aba:sec1}
In Deep Inelastic Scattering (DIS), the structure of the nucleon is probed by the interaction of
a high energy lepton and a target nucleon, via the 
exchange of a virtual boson. If at least one of the 
produced hadrons is detected in coincidence with the scattered lepton, 
the reaction is called Semi-Inclusive Deep Inelastic Scattering (SIDIS):
\begin{equation}
l({\bf k})\,+\,N({\bf P})\, \rightarrow \, l'({\bf k}')\, + \,h({\bf P}_h)\,\,+\,X({\bf P}_X),
\end{equation}
where $l$ ($l'$) is the incident (scattered) beam lepton, 
$N$ is the target nucleon, $h$ the detected hadron and $X$ 
the target remnant. The quantities in parentheses are
the corresponding four-momenta.

\begin{figure}
  \includegraphics[height=.15\textheight]{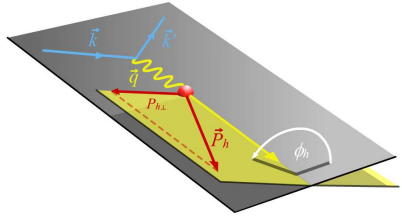}
  \caption{Definition of the azimuthal angle $\phi_h$ between the scattering plane 
(grey) and the hadron production 
plane (yellow).}
\label{fig:evento}
\end{figure}

If the cross section is unintegrated over the hadron momentum component 
transverse to the virtual photon direction $P_{h\perp}$ (Fig.~\ref{fig:evento}), 
an azimuthal dependence around the outgoing hadron direction exists~\cite{Bacchetta}:
\begin{equation}
\centering 
\begin{split}
\frac{d\sigma}{dx\,dy\,dz\,dP^2_{h\perp}\,d\phi_h}=
\frac{\alpha^2}{xyQ^2}(1+\frac{\gamma^2}{2x})
\{A(y)\,F_{UU,T}+B(y)\, F_{UU,L}+\\ C(y)\,\csm F_{UU}^{\csm}+B(y)\,\cstm F_{UU}^{\cstm}\},
\label{eq:noncol_csec}
\end{split}
\end{equation}
where $\phi_h$ is the azimuthal angle of the hadron plane around the virtual-photon direction 
(Fig.~\ref{fig:evento}). Here $Q^2$ and $y$ are respectively the negative squared 
four-momentum and the fraction of the lepton's energy transferred to the virtual photon, 
$x$ is the Bjorken scaling variable and $z$ is the virtual photon's 
fractional energy transferred to the produced hadron.
For the structure functions $F$, the subscript $UU$ denotes Unpolarized beam and Unpolarized target, 
$T$ ($L$) indicates the Transverse (Longitudinal) polarization of the virtual photon, 
$\alpha$ is the electromagnetic coupling constant, $\gamma=2Mx/Q$ with $M$ the target mass,
$A(y) \approx (1-y+1/2y^2)$, $B(y) \approx (1-y)$ and $C(y) \approx (2-y)\sqrt{1-y}$.

Two mechanisms are expected to give important contributions to 
the azimuthal dependence of the unpolarized cross section in the hadron
transverse momentum range accessed at HERMES:

\hspace{0.5cm}- the Cahn effect, a pure kinematic effect, 
generated by the non-zero intrinsic transverse motion of quarks, 
already pointed out by R.~Cahn in 1978~\cite{Cahn};

\hspace{0.5cm}- the Boer-Mulders effect, introduced more recently (1997) 
by D.~Boer and P.~J.~Mulders~\cite{BoerMuld}; 
this mechanism originates from a coupling between quark transverse 
momentum and quark transverse spin.

\subsection{The HERMES experiment}\label{aba:sec2}
The fixed target HERMES experiment was operated for more than $10$
years until $2007$ at the electron-positron storage ring of HERA at DESY. 
The HERMES spectrometer~\cite{Spect} was a forward angle instrument 
consisting of two symmetric halves (top, bottom) above and below the horizontal plane. 
It was characterized by very high efficiency ($98 \div 99\%$) in electron-hadron separation, 
provided by a transition radiation detector, a preshower 
scintillation counter and an electromagnetic calorimeter. 
In addition, a dual-radiator Ring-Imaging CHerenkov detector (RICH) provided 
good hadron-type identification for momenta above 2 GeV/c.

\section{Multi-dimensional unfolding and results}\label{aba:sec3}
The cross section azimuthal modulations 
can be measured via the $\la{\cos n\phi_h}\ra$-moments:
\begin{equation}\label{eq:moments}
\la {\cos n\phi_h}\ra \,=\,\frac{\int \cos n\phi_h\,d^5\sigma}{\int d^5\sigma},
\end{equation}
where $n=1,2$ and $\int d^5\sigma$ stands for 
$\int dx\,dy\,dz\,dP^2_{h\perp}\,d\phi_h\,\frac{d^5\sigma}{dx\,dy\,dz\,dP^2_{h\perp}\,d\phi_h}$.

The extraction of these cosine moments from data 
is challenging because they couple to a number of azimuthal 
modulations that are due to experimental sources, {\it e.g.} detector
geometrical acceptance and higher-order QED effects (radiative effects). 
Moreover, typically the event sample is binned 
only in one variable ($1$-dimensional analysis),
and integrated over the full range of all the other ones, while 
the structure functions $F$ used in equation~\ref{eq:noncol_csec} and the instrumental contributions
depend on all the kinematic variables $x$, $y$, $z$ and $P_{h\perp}$ simultaneously.
 
Therefore, in order to determine the cosine moments corrected 
for radiative and detector smearing, an unfolding procedure~\cite{Cowan} was used,
in which the event sample is binned simultaneously in all the relevant 
variables ({\it multi-dimensional analysis}\footnote{For a more 
detailed discussion about $1$- and multi-dimensional 
analysis see~\cite{Transv08}.}).

The unfolding algorithm is based on the relation between the 
unknown distribution of {\it Born} yields $B(j)$ and the distribution
of {\it measured} yields $X(i)$:
\begin{equation}\label{eq:unf}
\begin{split}
X(i)=\sum_{j=1}^{n_b}S&(i,j)B(j)+\beta(i).
\end{split}
\end{equation}
where $n_b$ is the total number of bins and $\beta(i)$ is a vector 
that contains the events smeared into the 
measured sample from outside the acceptance. The 
{\it Smearing matrix} $S(i,j)$ describes
the probability that an event originating from the {\it Born bin} $j$, 
corresponding to the original kinematics (free from experimental 
distortions), is actually observed in the {\it measured bin} $i$. 
Both the background $\beta(i)$ and the smearing matrix $S(i,j)$ 
are determined by a detailed Monte Carlo simulation of the 
experimental apparatus. 
\begin{figure}
\parbox{15cm}
{\centering\includegraphics[width=0.6\linewidth]{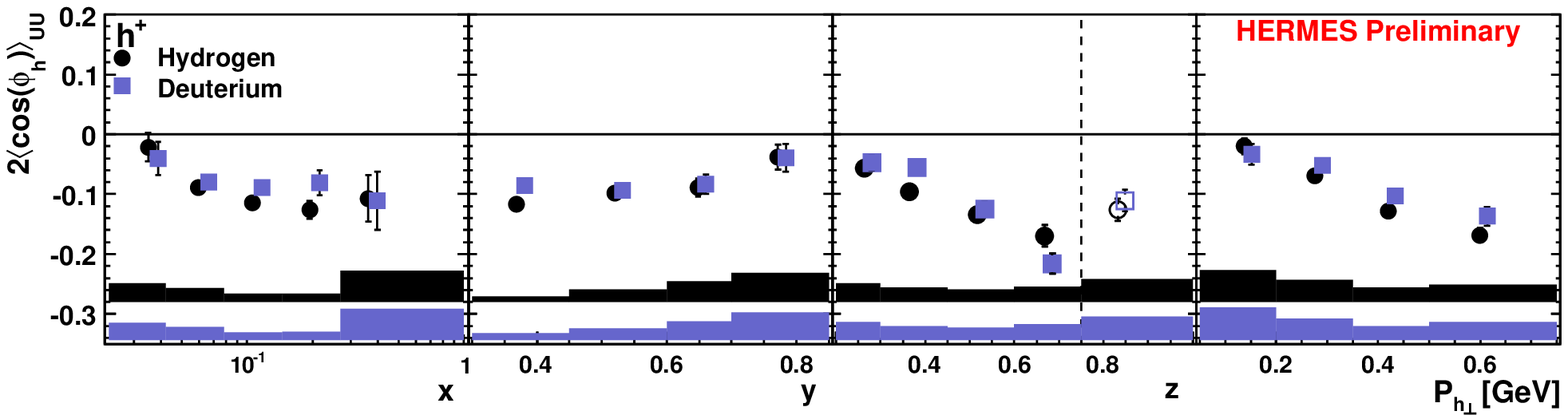}
\includegraphics[width=0.6\linewidth]{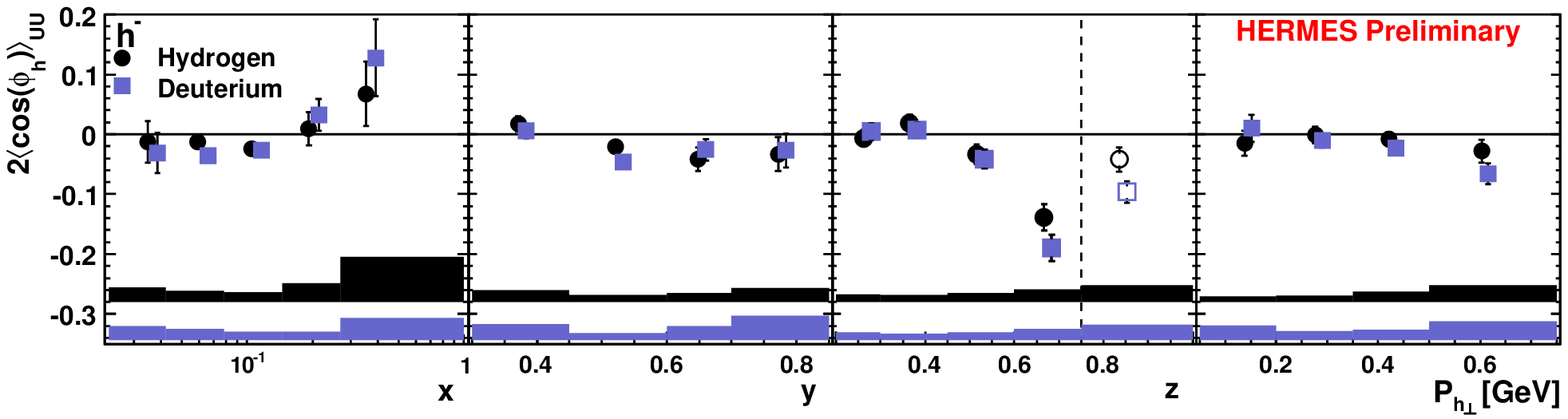}}
\caption{The $\cos\phi_h$ moments for positive (upper panel) and negative (lower panel) hadrons, 
extracted from hydrogen (circles) and deuterium (squares) data, shown as
projection versus the kinematic variables $x$, $y$, $z$ and $P_{h\perp}$.}
\label{fig:cosphi}
\end{figure}

Assuming a non-singular $S(i,j)$ matrix one obtains:
\begin{equation}
B(j)=\sum_{i=1}^{n_b} S^{-1}(j,i)\big[X(i)-\beta(i)\big].
\end{equation}
The extraction of cosine moments from the 
{\it Born} yields $B(j)$ can be performed by 
linear regression that takes into account 
the correlations between bins introduced by the smearing.
In this way one pair of moments can be obtained 
in each kinematic bin ($\cs$, $\cst$), which represents results
that are fully differential in all variables. 

The dependence of a moment  on a single variable can be obtained by projecting 
the fully differential result onto the variable under
study by weighting the moment in each bin $k$ with the corresponding 
unpolarized $4\pi$ cross section $\sigma^{4\pi}_k$ (defined by a Monte Carlo), 
for instance in case of the $\ell$-th $x$-bin:
$\la {\cos n\ph} \ra(x_\ell)=\sum_k \sigma^{4\pi}_k \la {\cos n\ph} \ra _k / \sum_k \sigma^{4\pi}_k$, 
where $k$ runs over all the $n_b$ bins corresponding to $x_\ell$.
\\

The $\cos \phi_h$ moments from  hydrogen and deuterium data are shown in 
figure~\ref{fig:cosphi} as projections versus the relevant kinematic variables
for both hadron charges.
Both hydrogen and  deuterium data show similar behavior: 
the $\cs$ moments are found to be sizable and negative for positive hadrons. 
The signal increases with $P_{h\perp}$ and with the hadron energy 
fraction $z$, except in the very high $z$ range, where the
partonic interpretation of the cross section is no longer valid\footnote{
The highest $z$-bin is plotted for completeness but it is not used for projecting moments onto the single variables.}.
The signal for the negative hadrons is significantly lower, but the dependence 
versus z and $P_{h\perp}$ exhibits similar features.

\begin{figure}
\parbox{15cm}
{\centering\includegraphics[width=0.6\linewidth]{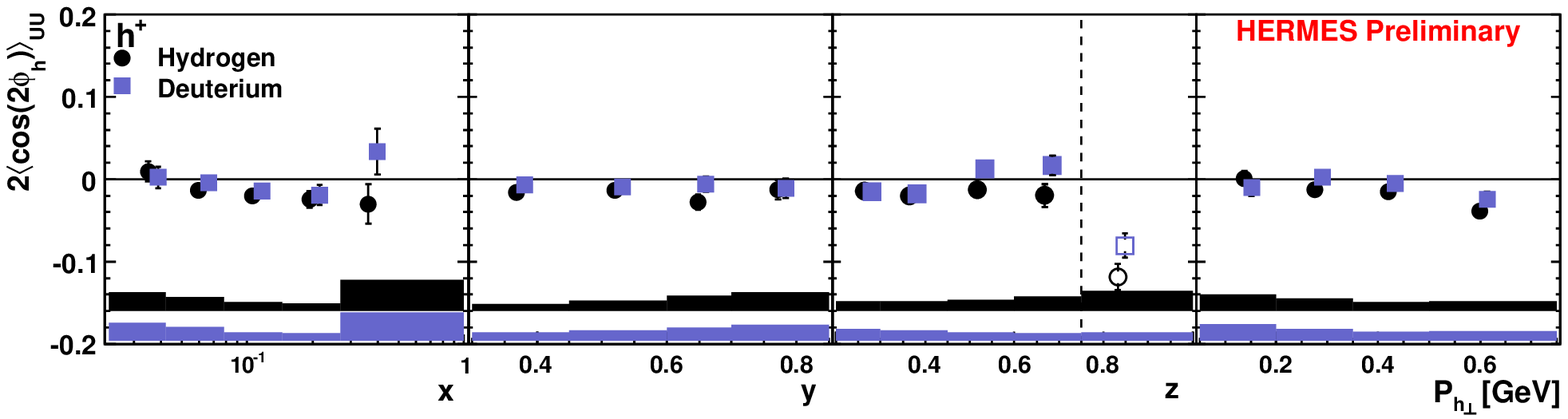}
\hspace{0.5cm}\includegraphics[width=0.6\linewidth]{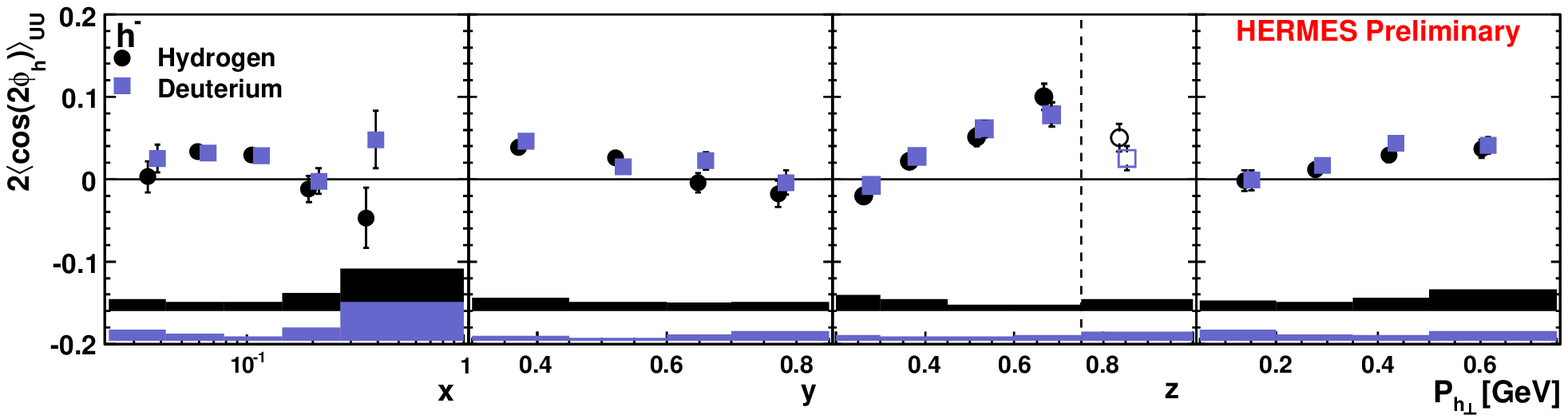}}
  \caption{The $\cos2\phi_h$ moments for positive (upper panel) and negative (lower panel) hadrons, 
extracted from hydrogen (circles) and deuterium (squares) data, shown as
projection versus the kinematic variables $x$, $y$, $z$ and $P_{h\perp}$.}
\label{fig:cos2phi}
\end{figure}

Figure~\ref{fig:cos2phi} shows the $\cos 2 \phi_h$ moments that
are found to be slightly negative for positive hadrons, and slightly positive
for negative hadrons.
Different results for positive and negative hadrons
are not unexpected because both experimental evidence and 
theoretical models predict opposite Boer-Mulders 
contributions for differently charged hadrons.

\end{document}